\documentclass[3p,times]{elsarticle}%
\usepackage{ecrc}
\usepackage{graphicx}
\usepackage{amsmath}
\usepackage{amsfonts}
\usepackage{amssymb}
\usepackage{subfig}
\usepackage{url}%
\providecommand{\U}[1]{\protect\rule{.1in}{.1in}}

\volume{00}
\firstpage{1}
\journalname{Procedia Computer Science}
\runauth{J. Mandel et al.}
\jid{procs}
\jnltitlelogo{Procedia Computer Science}
\begin{document}
\begin{frontmatter}
\title{Assimilation of Perimeter Data and Coupling with Fuel\\ Moisture in a Wildland Fire -- Atmosphere DDDAS}
\dochead{International Conference on Computational Science, ICCS 2012}
\author[ucd]{Jan Mandel\corref{cor1}}
\ead{Jan.Mandel@gmail.com}
\author[ucd]{Jonathan D. Beezley}
\author[ut]{Adam K. Kochanski}
\author[ucd]{Volodymyr Y. Kondratenko}
\author[ucd]{Minjeong Kim}
\address[ucd]{Department of Mathematical and
Statistical Sciences,\\ University of Colorado Denver, Denver, CO
80217-3364, USA}
\address[ut]{Department of Atmospheric Sciences,\\ University of Utah, Salt Lake City, UT
84112-0110, USA}
\cortext[cor1]{Corresponding author}
\begin{abstract}
We present a methodology to change the state of the Weather Research Forecasting (WRF) model
coupled with the fire spread code SFIRE, based on Rothermel's formula and the level set method,
and with a fuel moisture model.
The fire perimeter in the model changes in response to data while the model is running.
However, the atmosphere state takes time to develop in response to the forcing by the heat flux
from the fire. Therefore, an artificial fire history is created from an earlier fire perimeter
to the new perimeter, and replayed with the proper heat fluxes to allow the atmosphere state
to adjust. The method is an extension of an earlier method to start the coupled fire model
from a developed fire perimeter rather than an ignition point. The level set method can be also
used to identify parameters of the simulation, such as the fire spread rate.
The coupled model is available from \url{openwfm.org}, and it extends the WRF-Fire code
in WRF release.
\end{abstract}
\end{frontmatter}

\section{Introduction}

\begin{figure}[t]
\vspace*{-0.7in}
\par
\begin{center}
\includegraphics[width=5.5in]{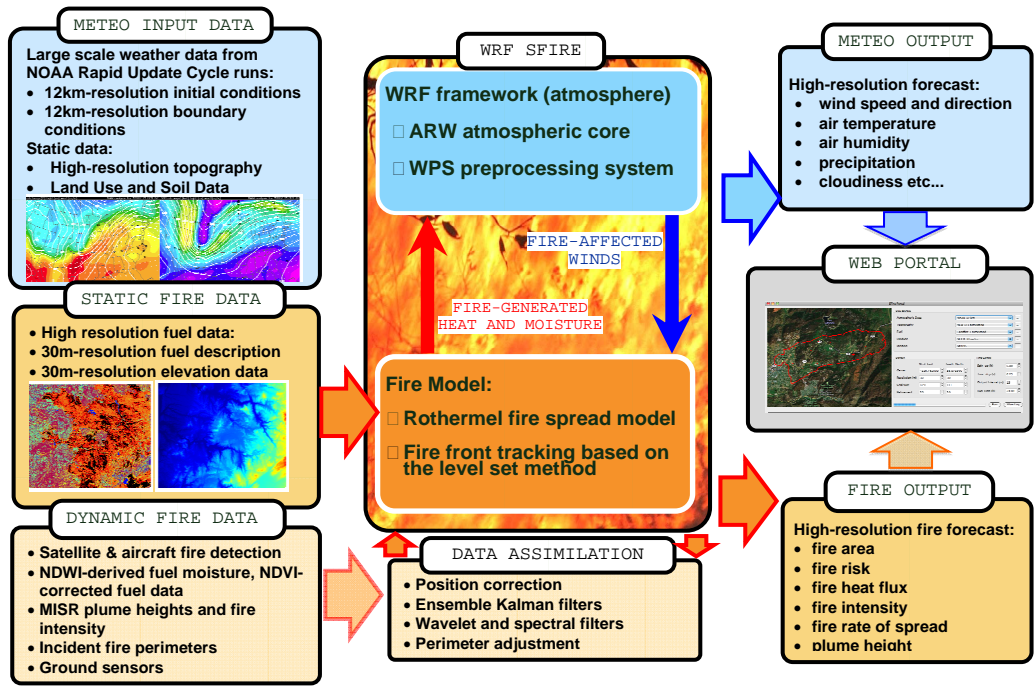}
\end{center}
\caption{Scheme of wildland fire DDDAS. }%
\label{fig:scheme}%
\end{figure}

This article reports on recent developments in building a Dynamic Data Driven
Application System (DDDAS) for wildland fire simulations
\cite{Mandel-2004-NDD,Douglas-2006-DVW-x,Beezley-2008-RDD}. A DDDAS\ is based
on the ability to incorporate data into an executing simulation
\cite{Darema-2004-DDD}. See Fig.~\ref{fig:scheme} for the overall scheme of
the DDDAS.

The paper is organized as follows. In Sec.~\ref{sec:assimilation-fires}, we we
review some existing approaches to data assimilation in simulations of
wildland fires. In Sec.~\ref{sec:model} and \ref{sec:replay}, we briefly
formulate the model and the principal idea of creating and replaying
artificial fire history from point ignition to a given perimeter, for
reference. Sec.~\ref{sec:create} considers several methods for the construction
of a level set function, needed for the replay, two from our previous work and
a new method, which is an extension of the reinitialization equation approach
known in level set methods. In Sec.~\ref{sec:assimilation-level}, we present a new
method how the level set functions constructed for two perimeters can be used
to create and replay an artificial fire history between the two perimeters,
and a new method that uses the two level set functions for an automatic
adjustment of the fire spread rate between the two perimeters.
Sec.~\ref{sec:moisture} describes a new moisture model coupled with the fire
and atmosphere model, and the possibilities for the assimilation of moisture
data. Finally, Sec.~\ref{sec:conclusion} is the conclusion.

\section{Data assimilation for wildland fires}

\label{sec:assimilation-fires} One way to incorporate data into an executing
simulation is by sequential statistical estimation, which takes all available
data to date into account, and is known in geosciences as data assimilation.
Data assimilation is a standard technique in numerical weather prediction, and
the ability to assimilate large amounts of real-time data is behind much of
the recent improvement of weather forecast skill \cite{Kalnay-2003-AMD}.
However, data assimilation for wildland fires poses unique challenges, and
classical data assimilation methods simply will not work
\cite{Beezley-2008-MEK,Johns-2008-TEK,Mandel-2009-DAW}. One of the reasons is
that in many other physical systems where standard methods work well, such as
pollution transport or atmospheric dynamics, unwanted perturbations tend to
dissipate over time; but, in a fire model, once a perturbation ignites an
unwanted fire, the fire will keep growing, and after few assimilation cycles,
everything burns. Another reason is that a fire as a coherent structure, needs
to be moved, started, or extinguished in response to the data, which requires
positional, Lagrangean correction; additive corrections of the values of the
physical fields are not very useful.

Data assimilation methods by sequential Monte-Carlo methods (SMC), also known
as particle filters, were developed in the literature for cell-based fire
models \cite{Bianchini-2006-IPM-x,Gu-2008-TAP}. They can handle
non-Gaussian distributions, but they are computationally very expensive,
because they require very large ensembles to cover a region of the state space
by random perturbations.
A suitable perturbation algorithm is the key to a successful application. The
perturbation methods used in wildland fire modeling range from random
modifications of the burn area \cite{Bianchini-2006-IPM-x} to genetic
algorithms, which evolve the shape of the fire by simulated evolution, where
the states with fire regions closer the the data are more likely to survive
\cite{Denham-2009-CSS-x}. While SMC methods with tens of thousands of
particles may be feasible for 2D cell models, with relatively small state
vectors, they are definitely out of question for a coupled atmosphere-fire
model. Methods based on the optimal statistical interpolation and the Kalman
filter (KF), such as the ensemble Kalman filter (EnKF), assume that the state
distribution is at least approximately Gaussian and they modify the state in
response to data \cite[p.~180]{Kalnay-2003-AMD} rather than rely on hitting
the right answer with random perturbations. Thus, KF-based mehods require much
smaller ensembles that SMC methods, but still in the range of 20-100 members
and easily many hundreds \cite{Evensen-2009-DAE}. However, because of the fine
resolution of the atmospheric model needed over large areas, and the
associated need for small time steps, the simulations are computationally very
demanding, and such ensembles are still out of question. FFT-based data
assimilation methods, which reduce data assimilation to efficient operations
with diagonal matrices \cite{Mandel-2010-FFT-x} and can
drastically reduce the required ensemble size, from hundreds to often just 5
or 10 members. However, using the Fourier basis is tantamount to the
assumption that the state covariance does not vary spatially
\cite{Berre-2000-ESM-x}. Wavelet estimation can combine the effectiveness of
spectral methods with an automatic treatment of spatial locality
\cite{Deckmyn-2005-WAR}. Wavelet diagonal approximations of the covariance
matrix \cite{Pannekoucke-2009-HCM} are of particular interest, as they allow
efficient evaluation of the EnKF formulas \cite{Beezley-2011-WEK}. 

\begin{figure}[t]
\begin{center}
\includegraphics[width=5.5in]{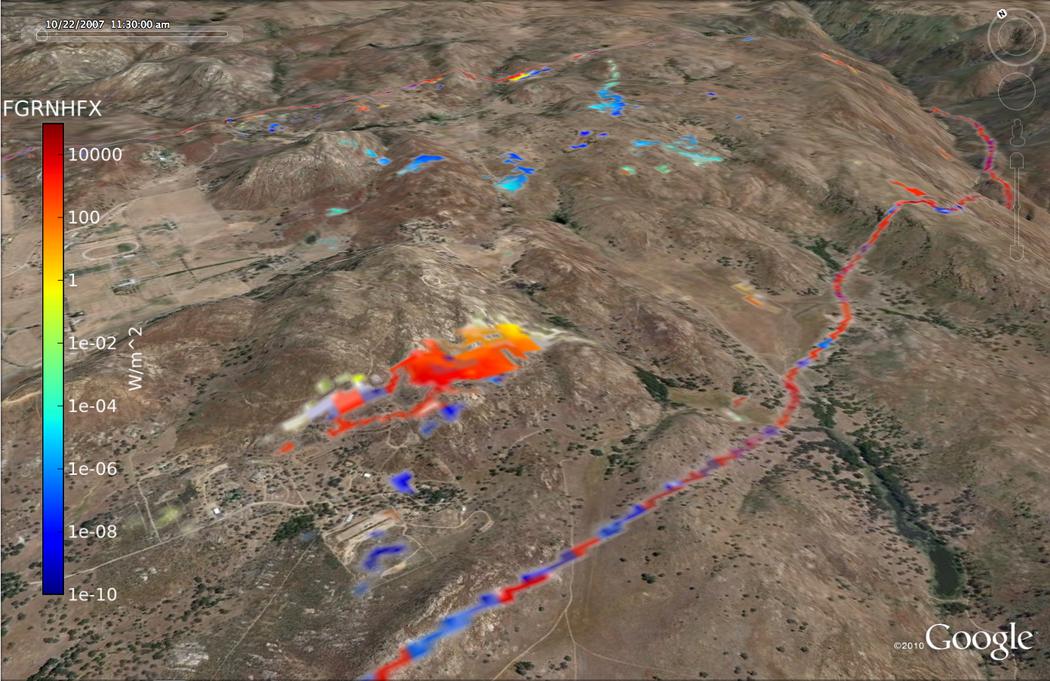}
\end{center}
\caption{Visualization in Google Earth client, 2007 Witch Fire. False color
shows fire heat flux, superimposed on the Earth surface. Patches of slower
fuels keep burning behind the fireline. Reproduced from \cite{Mandel-2011-WFM}%
. }%
\label{fig:google_earth}%
\end{figure}

Position correction methods, such as morphing \cite{Beezley-2008-MEK}, can
overcome the limitations of changing the state of the simulation by additive
corrections only. These method extend the state by a new variable containing a
deformation field, similarly as in optical flow methods
\cite{Marzban-2009-TSV} and extraction of the wind field from a sequence of
radar images \cite{Laroche-1994-VAM-x}. For other related position correction
methods, see, e.g., \cite{Ravela-2007-DAF}. Our morphing technique is
distinguished by replacing linear combinations of member states, which are at
the heart of, e.g., the EnKF, by intermediate states, which interpolate both
the magnitude and the position of coherent features, such as fires. Time
series of station observations could be handled by considering composite
states over several time steps. However, while morphing works successfully for
fire models \cite{Beezley-2008-MEK,Mandel-2009-DAW}, it changes the delicate
physical balance of the atmospheric equations and limits the possibility of
the treatment of the model as a black box. Even a simple linear transformation
to move and reshape the vortex in hurricane forecasting needs rebalancing of the
atmospheric variables from conservation equations \cite[p.~11]%
{Gopalakrishnan-2010-HWR}.

Therefore, an important problem in data assimilation for a coupled
atmosphere-fire model is how to adjust the atmosphere state when the state of
the fire model changes in response to data. The heat output of the fire is
concentrated in a narrow area with active combustion, therefore the fire
forcing on the atmosphere is highly localized. If the fire is just shifted, a
position correction alone can be successfull to some extent
\cite{Beezley-2008-MEK,Mandel-2009-DAW} because the relationship between the
changes in the atmosphere and in the fire is captured in the covariance of
their deformation fields. However, in general, the covariance does not contain
sufficient information and a spin-up is required to develop proper circulation
patterns for the changed fire forcing.



\section{The coupled atmosphere -- fire model}

\label{sec:model}

Over time, the wildland fire DDDAS has evolved from a simple
convection-reaction-diffusion equation exploratory model to test data
assimilation methodologies \cite{Mandel-2008-WFM} and the CAWFE model
\cite{Clark-1996-CAF-x,Clark-2004-DCA}, which couples the Clark-Hall
atmospheric model with fire spread implemented by tracers (Lagrangean
particles), to the currently used Weather Research Forecasting (WRF) mesoscale
atmospheric code \cite{Skamarock-2008-DAR} coupled with a spread model
implemented by the level set method \cite{Mandel-2009-DAW,Mandel-2011-CAF}.
The Clark-Hall model\ has many favorable properties, such as the ability to
handle refinement, but WRF is a supported community model, it can execute in
parallel, and has built-in export and import of state, which is essential for
data assimilation. Also, WRF supports data formats standard in geosciences.
The implementation by the level set method was chosen because the level set
function can be manipulated 
much more
easily than tracers. The coupled code is available from the Open Wildland Fire
Modeling Environment (OpenWFM) \cite{Mandel-2011-WFM} at \url{openwfm.org},
which contains also diagnostic and data processing utilities, including
visualization in Google Earth (Fig.~\ref{fig:google_earth}), which we first
proposed in \cite{Douglas-2006-DVW-x}. A subset of the SFIRE\ code was
released with WRF as WRF-Fire. The model is capable of running on a cluster
faster than real time with atmospheric resolution in tens of m, needed to
resolve the atmosphere-fire interaction, for a fire of size over $10$ km
\cite{Jordanov-2012-SHF}. 
See
\cite{Mandel-2009-DAW,Mandel-2011-CAF} for futher details and references.

The state variables of the fire model are the level set function, $\Phi$, the
time of ignition $T_{\mathrm{i}}$, and the fuel fraction remaining $F$, given
by their values on the nodes of the fire model mesh. At a given simulation
time $t$, the fire area is represented by the level set function $\Phi$ as the
set of all points $\left(  x,y\right)  $ where $\Phi\left(  t,x,y\right)
\leq0$. Since the level set function is interpolated linearly between nodes,
this allows a submesh representation of the fire area. In every time step of
the simulation, the level set function is advanced by one step of a
Runge-Kutta scheme for the level set equation%
\begin{equation}
\frac{d\Phi}{dt}=-R\left\vert \nabla\Phi\right\vert , \label{eq:level-set-eq}%
\end{equation}
where $R=R\left(  t,x,y\right)  $ is the fire rate of spread and $\left\vert
\cdot\right\vert $ is the Euclidean norm. The ignition time $T_{\mathrm{i}%
}=T_{\mathrm{i}}\left(  x,y\right)  $ is then computed for all newly ignited
nodes, and it satisfies the consistency condition%
\begin{equation}
\Phi\left(  t,x,y\right)  \leq0\Longleftrightarrow T_{\mathrm{i}}\left(
x,y\right)  \leq t, \label{eq:consistency}%
\end{equation}
where both inequalities express the condition that the location $\left(
x,y\right)  $ is burning at the time $t$.

The fire rate of spread $R$ is given by the Rothermel's formula
\cite{Rothermel-1972-MMP} as a function of the wind speed (at a height
dependent on the fuel) and the slope in the direction normal to the fireline.
From the level-set representation of the fireline at the time $t$ as
$\Phi\left(  t,x,y\right)  =0$, it follows by an easy calculus that the normal
direction is $\nabla\Phi/\left\vert \nabla\Phi\right\vert $, where $\left\vert
\cdot\right\vert $ is the Euclidean norm. Thus,%
\begin{equation}
R=R\left(  \mathbf{u}\cdot\frac{\nabla\Phi}{\left\vert \nabla\Phi\right\vert
}\right)  , \label{eq:spread-rate}%
\end{equation}
where $\mathbf{u}$ is the wind field.

Once the fuel starts burning, the remaining mass fraction $F=F(t,x,y)$ is
approximated by exponential decay,%
\begin{equation}
F\left(  t,x,y\right)  =\left\{
\begin{array}
[c]{cc}%
\exp\left(  -\frac{t-T_{\mathrm{i}}\left(  x,y\right)  }{T_{\mathrm{f}}\left(
x,y\right)  }\right)  , & t>T_{\mathrm{i}}\left(  x,y\right)  ,\\
1, & t\leq T_{\mathrm{i}}\left(  x,y\right)  ,
\end{array}
\right.  \label{eq:fuel}%
\end{equation}
where $T_{\mathrm{f}}$ is the fuel burn time, i.e., the number of seconds for
the fuel to burn down to $1/e\approx0.3689$ of the starting fuel fraction
$F=1$. The heat fluxes from the fire to the atmosphere are taken proportional
to the fuel burning rate, $\partial F\left(  t,x,y\right)  /\partial t$. The
proportionality constants are fuel coefficients. The heat fluxes from the fire
are inserted into the atmospheric model as forcing terms in differential
equations of the atmospheric model in a layer above the surface, with
exponential decay with altitude. This scheme is required because atmospheric
models with explicit timestepping, such as WRF, do not support flux boundary
conditions. The sensible heat flux is added to the time derivative of the
temperature, while the latent heat flux is added to the derivative of water
vapor concentration.

\section{Replaying artificial fire history}

\label{sec:replay}

The SFIRE code as presented in \cite{Mandel-2009-DAW,Mandel-2011-CAF} starts
from one or more ignition points. The release of the heat from the fire then
gradually establishes atmospheric circulation patterns and the fire evolves in
an interaction with the atmosphere. There is, however, a practical need to
start the simulation from an observed fire perimeter, and to modify the fire
perimeter in a running simulation, which presents a particular problem in a
coupled model. The atmospheric circulation due to the fire takes time to
develop and the heat release from the fire model needs to be gradual, or the
model will crash due to excessive vertical wind component.

Therefore, we have proposed creating and replaying an approximate fire
history, leading to the desired fire perimeter \cite{Kondratenko-2011-IFP}.
Replying the fire history allows for graduate release of the combustion heat
and allows the atmospheric circulation patterns due to the fire to develop.
The fire history is encoded as an array of ignition times $T_{\mathrm{i}%
}\left(  x,y\right)  $, prescribed at all fire mesh nodes. To replay the fire,
the numerical scheme for advancing $\Phi$ is suspended, and instead the level
set function is set to%
\begin{equation}
\Phi\left(  t,x,y\right)  =T_{\mathrm{i}}\left(  x,y\right)  -t.
\label{eq:fire-history-level-set-function}%
\end{equation}
The fuel decay (\ref{eq:fuel}) is then computed from $T_{\mathrm{i}}$, and the
resulting heat fluxes are inserted into the atmosphere. After the end of the
replay period is reached, the numerical scheme of the level set method takes over.

\section{Creating a level set function from a given fire perimeter}

\label{sec:create}

\begin{figure}[t]
\begin{center}
\includegraphics[width=3in]{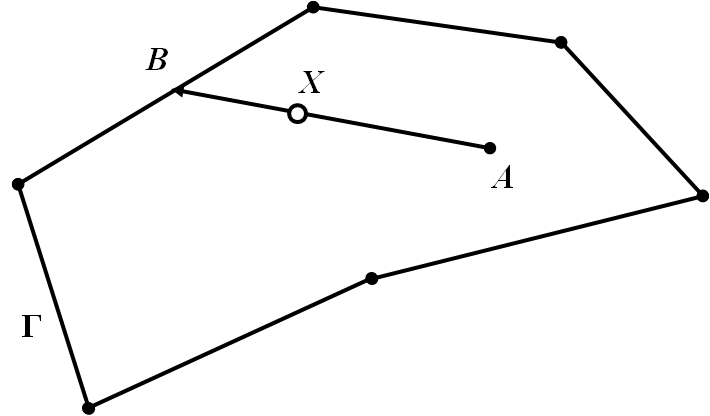}
\end{center}
\caption{Creating artificial time history by proportions. The ignition times
at the ignition point $A$ and the fire perimeter $\Gamma$ are interpolated
linearly along the segment between $A$ and a point $B$ on $\Gamma$ to a mesh
point $X$.}%
\label{fig:proportional}%
\end{figure}

\begin{figure}[t]
\begin{center}%
\begin{tabular}
[c]{cc}%
\hspace*{-0.3in} \includegraphics[height=2.5in]{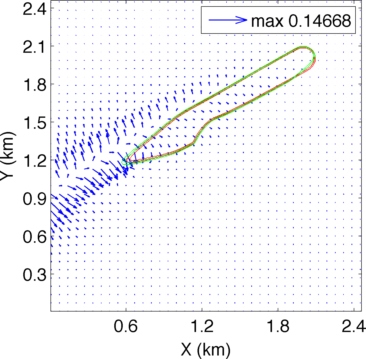} &
\hspace*{-0.3in} \includegraphics[height=2.5in]{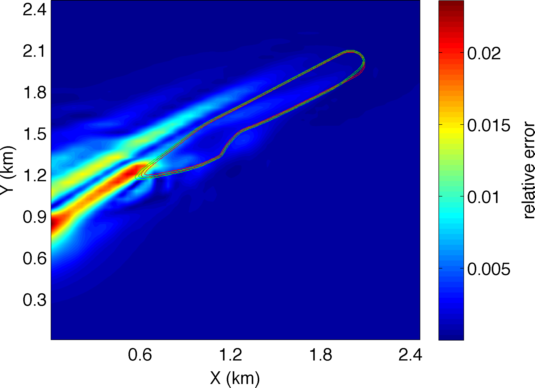}\\
(a) The difference in the horizontal wind vector (m/s). & (b) The relative
difference in the wind speed.
\end{tabular}
\end{center}
\caption{The difference in the horizontal wind field at $6.1$m between a
simulation propagated naturally from a point and another one advanced
artificially. The first simulation was ignited from a point in the northeast
corner of the domain, and the fire perimeter was recorded after $40$ minutes.
This perimeter and ignition location were used to generate an artificial
history for the first $40$ minutes, which was replayed in the second
simulation. Both simulations were then allowed to advance another $28$
minutes. Reproduced from \cite{Kondratenko-2011-IFP}. }%
\label{fig:coupled}%
\end{figure}

A fire perimeter is considered as a closed curve $\Gamma$, composed of linear
segments, and given as a sequence of the coordinates of the endpoints of the
segments. In practice, such geospatial data are often provided as a GIS
shapefile \cite{ESRI-1998-EST}, or encoded in a KML file, e.g., from LANDFIRE.

In \cite{Kondratenko-2011-IFP}, we have proposed a simple scheme for creating
an artificial fire history to be used in the fire replay scheme
(\ref{eq:fire-history-level-set-function}):\ given an ignition point and
ignition time at that point, approximate ignition times $T_{\mathrm{i}}$ at
the mesh points are established by linear interpolation between the ignition
point and the perimeter (Fig.~\ref{fig:proportional}). This simple method was
already shown to be successful in starting the model from the given perimeter
in a simple idealized case (Fig.~\ref{fig:coupled}), with the error in the
wind speed of only few \%. Extensions of the artificial history scheme will be
needed for domains which are not star-shaped with respect to the ignition
point. Running the fire propagation backwards in time to find an ignition
point is also a possibility, with an intriguing forensic potential
\cite{Kondratenko-2011-IFP}.

The ignition times $T_{\mathrm{i}}$ at locations outside of the given fire
perimeter are perhaps best thought of as what the ignition times at those
locations might be in future as the fire keeps burning.

Constructing a level set function from a perimeter is one of the basic tasks
in level set methods. Given a closed curve $\Gamma$, one wishes to construct a
function $L=L\left(  x,y\right)  $, such that
\begin{equation}
L>0\text{ outside of }\Gamma\text{,\quad}L<0\text{ inside of }\Gamma
\text{,\quad}L=0\text{ on }\Gamma. \label{eq:level-set-conditions}%
\end{equation}
In the application to perimeter ignition, one can then set at a fixed instant
$t$,%
\[
T_{\mathrm{i}}\left(  x,y\right)  =cL\left(  x,y\right)  +\left(  t-T\right)
,\quad\Phi\left(  t,x,y\right)  =L\left(  x,y\right)
\]
where $c$ is a scaling factor, and proceed with the replay as described in
Section \ref{sec:replay}.

\begin{figure}[t]
\begin{center}
\includegraphics[width=3.5in]{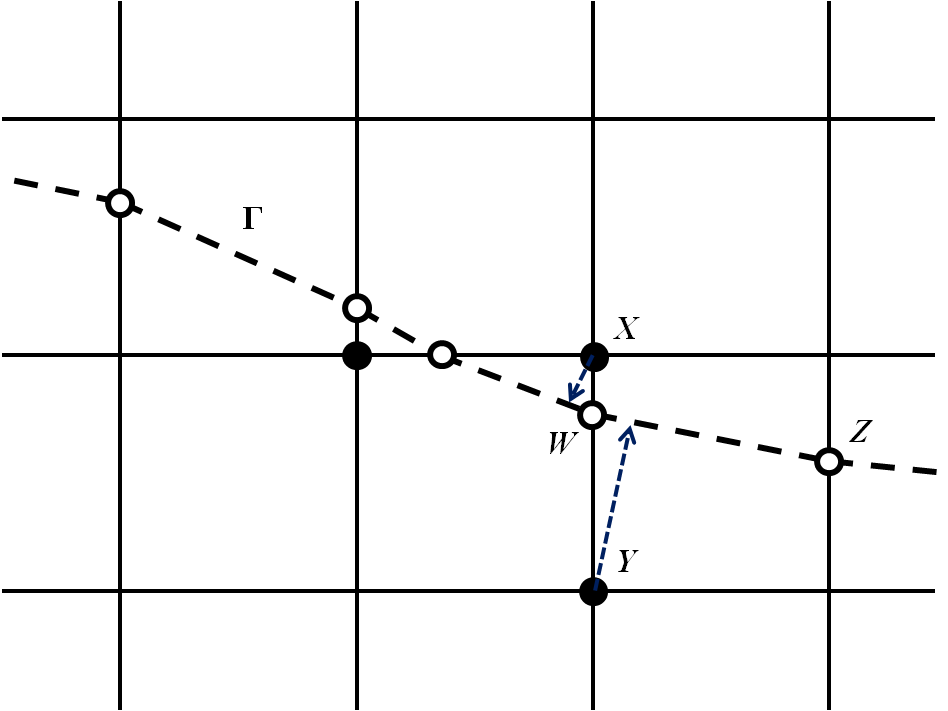}
\end{center}
\caption{A level set function linear on the line segments connecting the nodes
of the fire mesh cannot be defined at the nodes $X$ and $Y$ consistently as
the signed distance (\ref{eq:signed-dist}) from the interface $\Gamma$. The
distance of the point $X$ from $\Gamma$ does not depend on the location of the
point $Z$, while the distance of $Y$ does; yet the values of the level set
function at $X$ and $Y$ are linear along the segment $XY$ and so fixed by the
ratio of their distances from $W$.}%
\label{fig:corner}%
\end{figure}

One commonly used level set function is the signed distance from the given
closed curve $\Gamma$,
\begin{equation}
L\left(  x,y\right)  =\pm\operatorname*{dist}\left(  \left(  x,y\right)
,\Gamma\right)  , \label{eq:signed-dist}%
\end{equation}
where the sign is taken to be negative inside the region limited by $\Gamma$
and positive outside \cite{Osher-2003-LSM}, and $\operatorname*{dist}$ stands
for the Euclidean distance. Surprisingly, such function cannot be defined
consistently once the problem is discretized. Consider a level set function
$L$ that is given by its values on the corners of grid cells, interpolated
linearly along the grid lines, and $\Gamma$ given by its intersection with the
grid lines (Fig.~\ref{fig:corner}). Then, \emph{the ratio of the values of
}$L$\emph{ at two neighboring mesh corners on the opposite sides of }$\Gamma
$\emph{ is fixed by the requirement that }$L$\emph{ is linear between the two
corners}. In particular, it is not possible in general to define $L$ as the
signed distance (\ref{eq:signed-dist}). For example, in Fig.~\ref{fig:corner},
the ratio $L\left(  X\right)  /L\left(  Y\right)  $ is fixed and $L\left(
X\right)  $ does not depend on $Z$, while $L\left(  Y\right)  $ does.

One possibility is simply define the values of $L$ next to $\Gamma$ by the
signed distance, and forget about the exact representation of $\Gamma$ as
$L=0.$ Instead, in \cite{Mandel-CLF-2006}, we have proposed to find the values
of $L$ next to $\Gamma$ by least squares. Denoting by $u$ the vector of the
values $L$ next to $\Gamma$, it is easy to see that $u$ satisfies a
homogeneous system of linear equations of the form $Bu=0$ with at most two
nonzeros per rows, and each row corresponding to an edge on the mesh that is
intersected by \emph{ }$\Gamma$, as the edge $XY$. We can then find a suitable
$u$ minimizing $\left\Vert u-d\right\Vert ^{2}$ subject to $Bu=0$, where $d$
are the signed distances (\ref{eq:signed-dist}). Once the values of $L$ near
$\Gamma$ are found, one can extend $L$ to the whole domain as the distance
function by the Fast Marching Method (FMM) \cite{Sethian-1999-LSM}, or by a
simpler and less accurate approximate method suggested in
\cite{Mandel-CLF-2006}.

A better method can be obtained by taking the spread rate into account. The
level set function $L$ is a solution of the
Hamilton-Jacobi equation%

\[
R\left\vert \triangledown L\right\vert =1,\quad L=0\text{ on }\Gamma.
\]
which can be found by solving the reinitialization equation \cite[Eq.~(7.4)]%
{Osher-2003-LSM}%

\begin{equation}
\frac{\partial L}{\partial t}=\pm\left(  1-R\left\vert \triangledown
L\right\vert \right)  \label{eq:reinitialization}%
\end{equation}
where the sign is taken positive outside of $\Gamma$ and negative inside.
Equation (\ref{eq:reinitialization}) is solved by upwinding formulas moving
away from $\Gamma$ and starting from the values of $L$ on the other side of
$\Gamma$. Alternating the solution process between the outside and the inside
of $\Gamma$, the values of $L$ on the two sides of $\Gamma$ \textquotedblleft
balance out and a steady-state signed distance function is
obtained\textquotedblright\ \cite[p.~66]{Osher-2003-LSM}.

The situation here is more complicated, because the spread rate $R$ depends on
the level set function $L$ following (\ref{eq:spread-rate}). Hence, we freeze
$L$ inside $R$ and use successive approximations of the form%
\[
\frac{\partial L_{k+1}}{\partial t}=\pm\left(  1-R\left(  \mathbf{u}\cdot
\frac{\triangledown L_{k}}{\left\vert \triangledown L_{k}\right\vert }\right)
\left\vert \triangledown L_{k+1}\right\vert \right)  .
\]

\section{Data assimilation for the level set fire spread model}

\label{sec:assimilation-level}

Creating a level set function as in \cite{Kondratenko-2011-IFP} and in
Sec.~\ref{sec:create} allows for starting the coupled model from a given fire
perimeter instead of an ignition point. However, a more general approach is
needed for data assimilation. Suppose the fire perimeter in the simulation is
$\Gamma_{1}$ at time $t_{1}$. Then at time $t_{2}>t_{1}$, the fire evolves to
fire perimeter $\Gamma_{2}$. However, data is assimilated, changing the state
of the fire model and resulting in a different fire model state with perimeter
$\Gamma_{\mathrm{a}}$. First we construct level set functions $L_{1}$, $L_{2}%
$, and $L_{\mathrm{a}}$ for the perimeters $\Gamma_{1}$, $\Gamma_{2}$, and
$\Gamma_{\mathrm{a}}$, respectively, satisfying (\ref{eq:level-set-conditions}%
). We assume that all three level set functions are created using the same
method. \emph{The resulting approximate formulas will be exact in the case of
1D propagation with the level set functions linear and having the same slope}.
They will be used point-wise as an approximation otherwise. To emphasize the
point-wise application, we write out the arguments $\left(  x,y\right)  $ when present.

\subsection{Modifying the fire perimeter dynamically}

The state of the atmosphere will no longer match the state of the fire model
with the perimeter $\Gamma_{\mathrm{a}}$, and we need to make up the evolution
of the atmosphere as the fire progresses from the perimeter $\Gamma_{1}$ to
the perimeter $\Gamma_{\mathrm{a}}$. Since $\Gamma_{1}$ is completely
contained inside $\Gamma_{\mathrm{a}}$, in the region between $\Gamma_{1}$ and
$\Gamma_{\mathrm{a}}$, we have $L_{1}>0$ and $L_{\mathrm{a}}<0$. The function%
\begin{equation}
f_{1,a}\left(  x,y\right)  =\frac{L_{1}\left(  x,y\right)  }{L_{1}\left(
x,y\right)  -L_{\mathrm{a}}\left(  x,y\right)  } \label{eq:f1a}%
\end{equation}
then satisfies%
\[
f_{1,a}=0\text{ on }\Gamma_{1},\text{\quad}0<f_{1,a}<1\text{ between }%
\Gamma_{1}\text{ and }\Gamma_{a},\quad f_{1,a}=1\text{ on }\Gamma_{2}\text{.}%
\]

We can then use the function $f_{1,a}$ to create artificial ignition times by%
\[
T_{\mathrm{i}}\left(  x,y\right)  =t_{1}+\frac{L_{1}\left(  x,y\right)
}{L_{1}\left(  x,y\right)  -L_{\mathrm{a}}\left(  x,y\right)  }\left(
t_{2}-t_{1}\right)
\]
which interpolates between the perimeters $\Gamma_{1}$ and $\Gamma
_{\mathrm{a}}$, and replay the fire history to release the heat into the
atmosphere gradually, as in Sec.~\ref{sec:replay}.

\subsection{Dynamic estimate of fire spread rate}

\label{sec:estimate}A common source of errors in fire modeling is incorrect
spread rate.\ The level set function construction here can be used to adjust
the spread rate as well. Define $f_{1,2}$ similarly to (\ref{eq:f1a}),%
\begin{equation}
f_{1,2}\left(  x,y\right)  =\frac{L_{1}\left(  x,y\right)  }{L_{1}\left(
x,y\right)  -L_{2}\left(  x,y\right)  }. \label{eq:f12}%
\end{equation}
We now use a simple argument of proportions. Assume for the moment 1D fire
propagation in one direction and that $f_{1,a}$ and $f_{1,2}$ are linear. Then
$\Gamma_{1}$, $\Gamma_{2}$, and $\Gamma_{\mathrm{a}}$ are points on the real
line. and the spread rates of the simulated fire and the spread rate after the
data assimilation are, respectively,
\[
R=\frac{\Gamma_{2}-\Gamma_{1}}{t_{2}-t_{1}},\text{\quad}R_{\mathrm{a}}%
=\frac{\Gamma_{\mathrm{a}}-\Gamma_{1}}{t_{2}-t_{1}}.
\]
However, since $f_{1,2}$ and $f_{1,a}$ are linear,%
\[
f_{1,2}\left(  x\right)  =\frac{x-\Gamma_{1}}{\Gamma_{2}-\Gamma_{1}%
},\text{\quad}f_{1,a}\left(  x\right)  =\frac{x-\Gamma_{1}}{\Gamma_{a}%
-\Gamma_{1}},
\]
which gives%
\begin{equation}
\frac{R_{\mathrm{a}}\left(  x\right)  }{R\left(  x\right)  }=\frac
{\Gamma_{\mathrm{a}}-\Gamma_{1}}{\Gamma_{2}-\Gamma_{1}}=\frac{f_{1,2}\left(
x\right)  }{f_{1,a}\left(  x\right)  },\text{\quad}x\neq\Gamma_{1}.
\label{eq:ratio-R}%
\end{equation}

Thus, using (\ref{eq:f1a}) and (\ref{eq:f12}), (\ref{eq:ratio-R}) suggests to
modify the given spread rate $R$ at a point $\left(  x,y\right)  $ to become
the spread rate $R_{a}$ after the data assimilation, given by%
\begin{equation}
R_{\mathrm{a}}\left(  x,y\right)  =\frac{L_{1}\left(  x,y\right)
-L_{a}\left(  x,y\right)  }{L_{1}\left(  x,y\right)  -L_{2}\left(  x,y\right)
}R\left(  x,y\right)  . \label{eq:rate-correct}%
\end{equation}

\section{Moisture model}

\label{sec:moisture}

Fire spread rate depends strongly on the moisture contents of the fuel. In
fact, the spread rate drops to zero when the moisture reaches the so-called
extinction value. For this reason, we have coupled the fire
spread model with a simple fuel moisture model integrated in SFIRE and run
independently at every point of the mesh. The temperature and the relative
humidity of the air (from the WRF atmosphere model) determine the fuel
equilibrium moisture contents $E$ \cite{Viney-1991-RFF}, and
the actual moisture contents $m=m\left(  t\right)  $ is then modeled by the
standard time-lag equation%
\begin{equation}
\frac{dm}{dt}=\frac{E-m}{T_{\mathrm{d}}}, \label{eq:moisture-ode}%
\end{equation}
where $T_{\mathrm{d}}$ is the drying lag time. We use the standard model with
the fuel consisting of components with $1$, $10$, and $100$ hour lag time,
with the proportions given by the fuel category \cite{Scott-2005-SFB}, and the
moisture is tracked in each component separately. 
During rain,
the equilibrium moisture $E$ is replaced by the saturation moisture contents
$S$, and the equation is modified to achieve the rain-wetting lag time $T_{\mathrm{r}%
}$ only asymptotically for heavy rain,%
\begin{equation}
\frac{dm}{dt}=\frac{S-m}{T_{\mathrm{r}}}\left(  1-\exp\left(  -\frac{r-r_0}{r_{\mathrm{s}}%
}\right)  \right), \text{ if } r>r_0, \label{eq:rain-ode}
\end{equation}
where $r$ is the rain intensity, $r_0$ is the threshold rain intensity below which no perceptible
wetting occurs, and $r_\mathrm{s}$ is the saturation rain intensity,
at which $1-1/e\approx63\%$ of the maximal rain-wetting rate is achieved. The
coefficients can be calibrated to achieve a similar behavior as accepted
empirical models \cite{Fosberg-1971-DHT,VanWagner-1985-EFP}. See
\cite{Nelson-2000-PDC,Weise-2005-CTM} for other, much more
sophisticated models.
If moisture measurements are available, they can be ingested in the model
(\ref{eq:moisture-ode}, \ref{eq:rain-ode}) by a fast and cheap Kalman filter in one variable, run
at each point independently. 

\section{Conclusion}

\label{sec:conclusion}

We have presented new techniques to assimilate the perimeter data at two
different times into a coupled atmosphere-fire model, a new method to estimate
the adjustment of the model spread rate between the perimeters towards the
data, and a new coupling of the atmosphere-fire model with a third model, a
simple time-lag model of fuel moisture. Implementation of the data assimilation is in progress.
The moisture model is currently included in the code download
and will be treated in more detail elsewhere.

\section*{Acknowledgements}

This work was partially supported by the National Science Foundation under
grant EGS-0835579, by the National Institute of Standards and Technology
Fire Research Grants Program grant 60NANB7D6144, and by the Israel Department
of Homeland Security through Weather It Is, Inc. The help and encouragement provided by
Barry H. Lynn and Guy Kelman from Weather It Is, Inc. is appreciated.

\section*{References}

\bibliographystyle{elsarticle-num}
\bibliography{../../references/geo,../../references/other}

\begin{thebibliography}{10}
\expandafter\ifx\csname url\endcsname\relax
  \def\url#1{\texttt{#1}}\fi
\expandafter\ifx\csname urlprefix\endcsname\relax\def\urlprefix{URL }\fi
\expandafter\ifx\csname href\endcsname\relax
  \def\href#1#2{#2} \def\path#1{#1}\fi

\bibitem{Mandel-2004-NDD}
J.~Mandel, M.~Chen, L.~P. Franca, C.~Johns, A.~Puhalskii, J.~L. Coen, C.~C.
  Douglas, R.~Kremens, A.~Vodacek, W.~Zhao, A note on dynamic data driven
  wildfire modeling, in: M.~Bubak, G.~D. van Albada, P.~M.~A. Sloot, J.~J.
  Dongarra (Eds.), Computational Science -- ICCS 2004, Vol. 3038 of Lecture
  Notes in Computer Science, Springer, 2004, pp. 725--731.
\newblock \href {http://dx.doi.org/10.1007/b97989} {\path{doi:10.1007/b97989}}.

\bibitem{Douglas-2006-DVW-x}
C.~C. Douglas, J.~D. Beezley, J.~Coen, D.~Li, W.~Li, A.~K. Mandel, J.~Mandel,
  G.~Qin, A.~Vodacek, Demonstrating the validity of a wildfire {DDDAS}, in:
  V.~N. Alexandrov, D.~G. van Albada, P.~M.~A. Sloot, J.~Dongarra (Eds.),
  Computational Science -- ICCS 2006, Vol. 3993 of Lecture Notes in Computer
  Science, Springer, 2006, pp. 522--529.
\newblock \href {http://dx.doi.org/10.1007/11758532_69}
  {\path{doi:10.1007/11758532_69}}.

\bibitem{Beezley-2008-RDD}
J.~Beezley, S.~Chakraborty, J.~Coen, C.~Douglas, J.~Mandel, A.~Vodacek,
  Z.~Wang, Real-time data driven wildland fire modeling, in: M.~Bubak, G.~van
  Albada, J.~Dongarra, P.~Sloot (Eds.), Computational Science -- ICCS 2008,
  Vol. 5103 of Lecture Notes in Computer Science, Springer, 2008, pp. 46--53.
\newblock \href {http://dx.doi.org/10.1007/978-3-540-69389-5_7}
  {\path{doi:10.1007/978-3-540-69389-5_7}}.

\bibitem{Darema-2004-DDD}
F.~Darema, Dynamic data driven applications systems: A new paradigm for
  application simulations and measurements, in: M.~Bubak, G.~van Albada,
  P.~Sloot, J.~Dongarra (Eds.), Computational Science -- ICCS 2004, Vol. 3038
  of Lecture Notes in Computer Science, Springer, 2004, pp. 662--669.
\newblock \href {http://dx.doi.org/10.1007/978-3-540-24688-6_86}
  {\path{doi:10.1007/978-3-540-24688-6_86}}.

\bibitem{Kalnay-2003-AMD}
E.~Kalnay, Atmospheric Modeling, Data Assimilation and Predictability,
  Cambridge University Press, 2003.

\bibitem{Beezley-2008-MEK}
J.~D. Beezley, J.~Mandel, Morphing ensemble {K}alman filters, Tellus 60A (2008)
  131--140.
\newblock \href {http://dx.doi.org/10.1111/j.1600-0870.2007.00275.x}
  {\path{doi:10.1111/j.1600-0870.2007.00275.x}}.

\bibitem{Johns-2008-TEK}
C.~J. Johns, J.~Mandel, A two-stage ensemble {K}alman filter for smooth data
  assimilation, {E}nvironmental and Ecological Statistics 15 (2008) 101--110.
\newblock \href {http://dx.doi.org/10.1007/s10651-007-0033-0}
  {\path{doi:10.1007/s10651-007-0033-0}}.

\bibitem{Mandel-2009-DAW}
J.~Mandel, J.~D. Beezley, J.~L. Coen, M.~Kim, Data assimilation for wildland
  fires: Ensemble {K}alman filters in coupled atmosphere-surface models, IEEE
  Control Systems Magazine 29~(3) (2009) 47--65.
\newblock \href {http://dx.doi.org/10.1109/MCS.2009.932224}
  {\path{doi:10.1109/MCS.2009.932224}}.

\bibitem{Bianchini-2006-IPM-x}
G.~Bianchini, A.~Cort\'{e}s, T.~Margalef, E.~Luque, Improved prediction methods
  for wildfires using high performance computing: A comparison, in:
  V.~Alexandrov, G.~van Albada, P.~Sloot, J.~Dongarra (Eds.), Computational
  Science -- ICCS 2006, Vol. 3991 of Lecture Notes in Computer Science,
  Springer, 2006, pp. 539--546.
\newblock \href {http://dx.doi.org/10.1007/11758501_73}
  {\path{doi:10.1007/11758501_73}}.

\bibitem{Gu-2008-TAP}
F.~Gu, X.~Hu, Towards applications of particle filters in wildfire spread
  simulation, in: WSC '08: Proceedings of the 40th Conference on Winter
  Simulation, IEEE, 2008, pp. 2852--2860.
\newblock \href {http://dx.doi.org/10.1109/WSC.2008.4736406}
  {\path{doi:10.1109/WSC.2008.4736406}}.

\bibitem{Denham-2009-CSS-x}
M.~Denham, A.~Cort{\'e}s, T.~Margalef, Computational steering strategy to
  calibrate input variables in a dynamic data driven genetic algorithm for
  forest fire spread prediction, in: Computational Science--ICCS 2009, Vol.
  5545 of Lecture Notes in Computer Science, Springer, 2009, pp. 479--488.
\newblock \href {http://dx.doi.org/10.1007/978-3-642-01973-9_54}
  {\path{doi:10.1007/978-3-642-01973-9_54}}.

\bibitem{Evensen-2009-DAE}
G.~Evensen, Data Assimilation: The Ensemble {K}alman Filter, 2nd Edition,
  Springer, 2009.
\newblock \href {http://dx.doi.org/10.1007/978-3-642-03711-5}
  {\path{doi:10.1007/978-3-642-03711-5}}.

\bibitem{Mandel-2010-FFT-x}
J.~Mandel, J.~D. Beezley, V.~Y. Kondratenko, Fast {F}ourier transform ensemble
  {K}alman filter with application to a coupled atmosphere-wildland fire model,
  in: A.~M. Gil-Lafuente, J.~M. Merigo (Eds.), Computational Intelligence in
  Business and Economics, Proceedings of MS'10, World Scientific, 2010, pp.
  777--784, available as arXiv:1001.1588.
\newblock \href {http://dx.doi.org/10.1142/9789814324441_0089}
  {\path{doi:10.1142/9789814324441_0089}}.

\bibitem{Berre-2000-ESM-x}
L.~Berre, {Estimation of synoptic and mesoscale forecast error covariances in a
  limited-area model}, {Monthly Weather Review} {128}~({3}) ({2000})
  {644--667}.
\newblock \href
  {http://dx.doi.org/10.1175/1520-0493(2000)128<0644:EOSAMF>2.0.CO;2}
  {\path{doi:10.1175/1520-0493(2000)128<0644:EOSAMF>2.0.CO;2}}.

\bibitem{Deckmyn-2005-WAR}
A.~Deckmyn, L.~Berre, A wavelet approach to representing background error
  covariances in a limited-area model, Monthly Weather Review 133~(5) (2005)
  1279--1294.
\newblock \href {http://dx.doi.org/10.1175/MWR2929.1}
  {\path{doi:10.1175/MWR2929.1}}.

\bibitem{Pannekoucke-2009-HCM}
O.~Pannekoucke, Heterogeneous correlation modeling based on the wavelet
  diagonal assumption and on the diffusion operator, Monthly Weather Review
  137~(9) (2009) 2995--3012.
\newblock \href {http://dx.doi.org/10.1175/2009MWR2783.1}
  {\path{doi:10.1175/2009MWR2783.1}}.

\bibitem{Beezley-2011-WEK}
J.~D. Beezley, J.~Mandel, L.~Cobb, Wavelet ensemble {K}alman filters, in:
  Proceedings of IEEE IDAACS'2011, Prague, September 2011, Vol.~2, IEEE, 2011,
  pp. 514--518.
\newblock \href {http://dx.doi.org/10.1109/IDAACS.2011.6072819}
  {\path{doi:10.1109/IDAACS.2011.6072819}}.

\bibitem{Mandel-2011-WFM}
J.~Mandel, J.~D. Beezley, A.~K. Kochanski, V.~Y. Kondratenko, L.~Zhang,
  E.~Anderson, J.~{Daniels II}, C.~T. Silva, C.~R. Johnson, A wildland fire
  modeling and visualization environment, Paper 6.4, Ninth Symposium on Fire
  and Forest Meteorology, Palm Springs, October 2011, available at
  \url{http://ams.confex.com/ams/9FIRE/webprogram/Paper192277.html}, retrieved
  December 2011 (2011).

\bibitem{Marzban-2009-TSV}
C.~Marzban, S.~Sandgathe, H.~Lyons, N.~Lederer, Three spatial verification
  techniques: Cluster analysis, variogram, and optical flow, {Weather and
  Forecasting} {24}~({6}) ({2009}) {1457--1471}.
\newblock \href {http://dx.doi.org/{10.1175/2009WAF2222261.1}}
  {\path{doi:{10.1175/2009WAF2222261.1}}}.

\bibitem{Laroche-1994-VAM-x}
S.~Laroche, I.~Zawadzki, {A variational analysis method for retrieval of
  three-dimensional wind field from single-Doppler radar data}, Journal of the
  Atmospheric Sciences 51~(18) (1994) 2664--2682.
\newblock \href
  {http://dx.doi.org/10.1175/1520-0469(1994)051<2664:AVAMFR>2.0.CO;2}
  {\path{doi:10.1175/1520-0469(1994)051<2664:AVAMFR>2.0.CO;2}}.

\bibitem{Ravela-2007-DAF}
S.~Ravela, K.~A. Emanuel, D.~McLaughlin, Data assimilation by field alignment,
  Physica D 230 (2007) 127--145.
\newblock \href {http://dx.doi.org/10.1016/j.physd.2006.09.035}
  {\path{doi:10.1016/j.physd.2006.09.035}}.

\bibitem{Gopalakrishnan-2010-HWR}
S.~Gopalakrishnan, Q.~Liu, T.~Marchok, D.~Sheinin, N.~Surgi, R.~Tuleya,
  R.~Yablonsky, X.~Zhang, Hurricane {Weather and Research and Forecasting
  (HWRF)} model scientific documentation, NOAA,
  \url{http://www.dtcenter.org/HurrWRF/users/docs/scientific_documents/HWRF_final_2-2_cm.pdf},
  retrieved October 2011 (2010).

\bibitem{Mandel-2008-WFM}
J.~Mandel, L.~S. Bennethum, J.~D. Beezley, J.~L. Coen, C.~C. Douglas, M.~Kim,
  A.~Vodacek, A wildland fire model with data assimilation, Mathematics and
  Computers in Simulation 79 (2008) 584--606.
\newblock \href {http://dx.doi.org/10.1016/j.matcom.2008.03.015}
  {\path{doi:10.1016/j.matcom.2008.03.015}}.

\bibitem{Clark-1996-CAF-x}
T.~L. Clark, M.~A. Jenkins, J.~Coen, D.~Packham, A coupled atmospheric-fire
  model: {C}onvective feedback on fire line dynamics, Journal of Applied
  Meteorolgy 35 (1996) 875--901.
\newblock \href
  {http://dx.doi.org/10.1175/1520-0450(1996)035<0875:ACAMCF>2.0.CO;2}
  {\path{doi:10.1175/1520-0450(1996)035<0875:ACAMCF>2.0.CO;2}}.

\bibitem{Clark-2004-DCA}
T.~L. Clark, J.~Coen, D.~Latham, Description of a coupled atmosphere-fire
  model, International Journal of Wildland Fire 13 (2004) 49--64.
\newblock \href {http://dx.doi.org/10.1071/WF03043}
  {\path{doi:10.1071/WF03043}}.

\bibitem{Skamarock-2008-DAR}
W.~C. Skamarock, J.~B. Klemp, J.~Dudhia, D.~O. Gill, D.~M. Barker, M.~G. Duda,
  X.-Y. Huang, W.~Wang, J.~G. Powers, A description of the {A}dvanced
  {R}esearch {WRF} version 3, NCAR Technical Note 475,
  \url{http://www.mmm.ucar.edu/wrf/users/docs/arw_v3.pdf}, retrieved December
  2011 (2008).

\bibitem{Mandel-2011-CAF}
J.~Mandel, J.~D. Beezley, A.~K. Kochanski, Coupled atmosphere-wildland fire
  modeling with {WRF} 3.3 and {SFIRE} 2011, Geoscientific Model Development 4
  (2011) 591--610.
\newblock \href {http://dx.doi.org/10.5194/gmd-4-591-2011}
  {\path{doi:10.5194/gmd-4-591-2011}}.

\bibitem{Jordanov-2012-SHF}
G.~Jordanov, J.~D. Beezley, N.~Dobrinkova, A.~K. Kochanski, J.~Mandel,
  B.~Soused\'{i}k, Simulation of the 2009 {H}armanli fire ({B}ulgaria), in:
  I.~Lirkov, S.~Margenov, J.~Wan\'{s}iewski (Eds.), 8th International
  Conference on Large-Scale Scientific Computations, Sozopol, Bulgaria, June
  6-10, 2011, Vol. 7116 of Lecture Notes in Computer Science, Springer, 2012,
  pp. 291--298, also available as arXiv:1106.4736.

\bibitem{Rothermel-1972-MMP}
R.~C. Rothermel, A mathematical model for predicting fire spread in wildland
  fires, {USDA Forest Service Research Paper INT-115},
  \url{http://www.treesearch.fs.fed.us/pubs/32533} (1972).

\bibitem{Kondratenko-2011-IFP}
V.~Y. Kondratenko, J.~D. Beezley, A.~K. Kochanski, J.~Mandel, Ignition from a
  fire perimeter in a {WRF} wildland fire model, Paper 9.6, 12th WRF Users'
  Workshop, National Center for Atmospheric Research, June 20-24, 2011,
  \url{http://www.mmm.ucar.edu/wrf/users/workshops/WS2011/WorkshopPapers.php},
  retrieved August 2011 (2011).

\bibitem{ESRI-1998-EST}
{ESRI} shapefile technical description, An ESRI White Paper, Environmental
  Systems Research Institute, Inc.,
  \url{http://www.esri.com/library/whitepapers/pdfs/shapefile.pdf}, retrieved
  January 2012 (1998).

\bibitem{Osher-2003-LSM}
S.~Osher, R.~Fedkiw, Level Set Methods and Dynamic Implicit Surfaces, Springer,
  New York, 2003.

\bibitem{Mandel-CLF-2006}
J.~Mandel, V.~Kulkarni, Construction of a level function for fireline data
  assimilation, CCM Technical Report 234, University of Colorado at Denver,
  \url{http://ccm.ucdenver.edu/reports/rep234.pdf} (June 2006).

\bibitem{Sethian-1999-LSM}
J.~A. Sethian, Level set methods and fast marching methods, 2nd Edition, Vol.~3
  of Cambridge Monographs on Applied and Computational Mathematics, Cambridge
  University Press, Cambridge, 1999.

\bibitem{Viney-1991-RFF}
N.~R. Viney, A review of fine fuel moisture modelling, International Journal of
  Wildland Fire 1~(4) (1991) 215--234.
\newblock \href {http://dx.doi.org/10.1071/WF9910215}
  {\path{doi:10.1071/WF9910215}}.

\bibitem{Scott-2005-SFB}
J.~H. Scott, R.~E. Burgan, Standard fire behavior fuel models: {A}
  comprehensive set for use with {R}othermel's surface fire spread model, Gen.
  Tech. Rep. RMRS-GTR-153. Fort Collins, CO: U.S. Department of Agriculture,
  Forest Service, Rocky Mountain Research Station,
  \url{http://www.fs.fed.us/rm/pubs/rmrs_gtr153.html} (2005).

\bibitem{Fosberg-1971-DHT}
M.~A. Fosberg, J.~E. Deeming, Derivation of the 1- and 10-hour timelag fuel
  moisture calculations for fire-danger rating, U.S. Forest Service Research
  Note RM-207, \url{http://hdl.handle.net/2027/umn.31951d02995763p} (1971).

\bibitem{VanWagner-1985-EFP}
C.~E. Van~Wagner, T.~L. Pickett, Equations and {FORTRAN} program for the
  {C}anadian forest fire weather index system, Canadian Forestry Service,
  Forestry Technical Report 33 (1985).

\bibitem{Nelson-2000-PDC}
R.~M. Nelson~Jr., Prediction of diurnal change in 10-h fuel stick moisture
  content, Canadian Journal of Forest Research 30~(7) (2000) 1071--1087.
\newblock \href {http://dx.doi.org/10.1139/x00-032}
  {\path{doi:10.1139/x00-032}}.

\bibitem{Weise-2005-CTM}
D.~R. Weise, F.~M. Fujioka, R.~M. Nelson~Jr., A comparison of three models of
  1-h time lag fuel moisture in {H}awaii, Agricultural and Forest Meteorology
  133 (2005) 28--39.
\newblock \href {http://dx.doi.org/10.1016/j.agrformet.2005.03.012}
  {\path{doi:10.1016/j.agrformet.2005.03.012}}.

\end{thebibliography}

\end{document}